%% file: anonyi.tex
\documentclass[letter]{ieice}
\usepackage[dvips]{graphicx}
\field{}
\title{On Answer Substitutions in Logic Programming}
\authorlist{
\authorentry{Keehang KWON}{m}{labelA}
}
\affiliate[labelA]{The authors are with Computer Eng., DongA University. email:khkwon@dau.ac.kr}
\received{2003}{1}{1}
\revised{2003}{1}{1}
\finalreceived{2003}{1}{1}

\makeatletter
\long\def\@makemyfntext#1{$^{\rm *}\ $ #1}

\long\def\@myfootnotetext#1{\insert\footins{\footnotesize
    \interlinepenalty\interfootnotelinepenalty 
    \splittopskip\footnotesep
    \splitmaxdepth \dp\strutbox \floatingpenalty \@MM
    \hsize\columnwidth \@parboxrestore
   \edef\@currentlabel{\csname p@footnote\endcsname\@thefnmark}\@makemyfntext
    {\rule{\z@}{\footnotesep}\ignorespaces
      #1\strut}}}

\def\myfootnotetext{\@ifnextchar
     [{\@xfootnotenext}{\xdef\@thefnmark{\thempfn}\@myfootnotetext}}
\makeatother

\input macros

\newcommand{\ball}{\all^o}
\newcommand{\bsome}{\some^o}

\renewcommand{\intp}{pv} 

\renewcommand{\intp}{pv} 

\newcommand{\muprolog}{{Prolog$^{\theta}$}}

\begin{document}
\maketitle
\begin{summary}
Answer substitutions play a central role in logic programming.
To support {\it selective} answer substitutions, we refine $\some x$ in goals 
 into two different versions:
the noisy version $\bsome x$ and the silent version $\some x$.
The main difference is that only the instantiation in $\bsome x$ will be recorded in the answer 
substitutions. Similarly for $\all x$.
In addition, we discuss the notion of don't-care constants and don't-know constants.
\end{summary}

\section{Introduction}\label{sec:intro}

The notion of answer substitutions  plays a key role in logic programming.
Unfortunately, the notion of answer substitutions in traditional logic languages
is unsatisfactory in the sense that
every instantiation is visible to the user.
For example, consider the program $phone(tom,cs,4450)$ which represents  that
$tom$ is a CS major and his phone number is 4450.
Now solving a query $\some x \some y phone(tom,x,y)$ 
 is a success with the answer substitutions  $x = cs, y = 4450$.
What if we want to see only Tom's phone number, but not his major?  
It is not possible to specify this in traditional logic programming.

To fix this problem, inspired by  
\cite{Jap03}, we  extend Prolog to  include the following new 
 formulas (called  noisy universal/existential
 quantifiers):\footnote{They are originally called choice quantifiers, denoted by 
$\sqcap,\sqcup$ in \cite{Jap03}.}

\begin{itemize}

\item $\ball x D$ where $D$ is a definite clause.

\item $\bsome x G$  where  $G$ is a goal formula in Prolog.
\end{itemize}
\noindent
In the above,  $\bsome x\ G$ is identical
to $\some x\ G$ with the difference 
that, in the former, the instantiation $t$ for $x$ is visible and recorded
 in the answer substitution, but not in the latter. Thus the latter is a $silent$ version of
the former that chooses a term $t$ for $x$ silently.
Similarly for  $\ball x D$.

Thus it is $\bsome x\ G$, rather than $\some x\ G$, that adequately captures the notion
of answer substitution. Furthermore, it is the notion of answer subsitution instead of proof theory
 that really counts
when dealing with logic programming.

Fortunately, in a simple setting such as Prolog, operational semantics  for these new quantifiers 
  can be easily obtained from $\all, \some$ 
by additionally recording each instantiation during execution\footnote{In a more
general setting where $\some x\ D, \bsome x\ D$ are allowed, 
there is a subtle yet critical difference between $\some$ and $\bsome$. 
That is, in $\bsome x$, $x$ must be instantiated to a ground term, while $x$ can be instantiated to an arbitrary term ( a new constant, for example) in $\some x$.}.
To be specific, 
 we adopt the following operational semantics for $\some$ and $\bsome$.

\begin{itemize}

\item $\intp(D, \some x G, nil )$ {\rm if}\ 
       $\intp(D, [t/x]G,\_)\ $

\item $\intp(D, \bsome x G, \lb x,t\rb)$ {\rm if}\  
    $\intp(D, [t/x]G,\_)\ $

\end{itemize}
\noindent where the third argument is used to record the  instantiation (if any) at the
current proof step.

This paper also deals with anonymous variables. Anonymous variables in a clause $D$
can be interpreted as $\all x D$, but not $\ball D$, while anonymous variables in a goal $G$
can be interpreted as $\some x G$.

    This paper proposes  \muprolog, an extension of Prolog with new quantifiers.
The remainder of this paper is structured as follows. We describe \muprolog\
  in
the next sections. 
Section~\ref{sec:conc} concludes the paper.

\section{The Language}\label{sec:logic}

The language is a version of Horn clauses
 with noisy quantifiers. 
It is described
by $G$- and $D$-formulas given by the syntax rules below:
\begin{exmple}
\>$G ::=$ \>   $A \sep  G \land G   \sep  \some x G  \sep \bsome x G$ \\   \\
\>$D ::=$ \>  $A  \sep G \supset A\ \sep \all x  D \sep  \ball x  D \sep D \land D $
\end{exmple}
\noindent
\newcommand{\sync}{up}
\newcommand{\async}{down}

In the rules above, 
$A$  represents an atomic formula.
A $D$-formula  is called a  Horn
 clause with noisy quantifiers. 

We describe a proof procedure based on uniform proofs \cite{MNPS91}. 
\newcommand{\bc}{bc}
 Note that proof search  alternates between 
two phases: the goal-reduction phase 
and the backchaining phase. 
In  the goal-reduction phase (denoted by $\intp(D,G,\theta)$), the machine tries to solve a goal $G$ from
a clause $D$  by simplifying $G$. 
If $G$ becomes an atom, the machine switches to the backchaining mode. 
In the backchaining mode (denoted by $bc(D_1,D,A,\theta)$), the machine tries 
to solve an atomic goal $A$ 
by first reducing a Horn clause $D_1$ to simpler forms  and then 
backchaining on the resulting 
 clause (via rule (1) and (2)). Below, \_ represents a don't-care value.

A proof of $\lb D,G\rb$ is a sequence of $P_1,\ldots,P_n$ where each $P_i$ is either
$bc(D_i,D,G_i,\theta_i)$ or $pv(D_i,G_i,\theta_i)$, with $P_n = pv(D,G,\theta_n)$.
Each $P_i$
is called a {\it proof step}. Each $\theta_i$ records an instantiation
(if any) performed at the $i$th proof step.

\begin{defn}\label{def:semantics}
Let $G$ be a goal and let $D$ be a program.
Then the notion of  proving $\lb D,G\rb$ with recording the  answer
substitution performed at the last proof step $\theta$ -- $\intp(D,G,\theta)$ -- 
 is defined as follows:

\begin{numberedlist}

\item  $\bc(A,D,A,nil)$. \% This is a success.

\item    $\bc((G_0\supset A),D,A, nil)$ if 
 $\intp(D, G_0,\_)$. \% backchaining

\item    $\bc(D_1 \land D_2),D,A, nil)$ if   $\bc(D_1,D,A, \theta_1)$ or  $\bc(D_2,D,A, \theta_2)$. \% $\land$-L

\item    $\bc(\all x D_1,D,A, nil )$ if   $\bc([t/x] D_1,D, A, \theta)$.
\%  $\lb x,t\rb$ is not recorded in  the answer substitution.

\item    $\bc(\ball x D_1,D,A, \lb x,t\rb )$ if   $\bc([t/x] D_1,D, A, \_)$.
\%  $\lb x,t\rb$ is  recorded in  the answer substitution.

\item    $\intp(D, A, nil)$ if   $\bc(D,D, A, \_)$. \% switch to backchaining mode


\item  $\intp(D, G_1 \land G_2, nil)$  if $\intp(D,G_1,\_)$  $and$
  $\intp(D,G_2,\_)$.


\item $\intp(D,\exists x G, nil)$  if $\intp(D,[t/x]G, \_)$
\%  $\lb x,t\rb$ is not included in the answer substituion.

\item $\intp(D,\bsome x   G, \lb x,t\rb)$  if $\intp(D,[t/x]G, \_)$
\%  $\lb x,t\rb$ is included in the answer substituion.

\end{numberedlist}
\end{defn}

\noindent  
These rules are straightforward to read.

Once a proof tree is built, execution is simple. It
simply displays its proof steps with the corresponding answer substitution
in a bottom-up manner.

\section{An Example}

 As an example, consider solving a query \\
 $\some x\ \bsome y\ phone(tom,x,y)$
 from the program $A$ of the form
$phone(tom,cs,4450)$.  Note that
this query can
be rewritten as $ \bsome y\ phone(tom,\_,y)$ using an anonymous variable.

Below is its proof with answer substitution $\lb y, 4450\rb$. \\

1. $bc(A,A,A,nil)$

2. $pv(A,A,nil)$

3. $pv(A,\bsome y\ phone(tom,cs,y), \lb y, 4450\rb)$

4. $pv(A,\some x \bsome y\ phone(tom,x,y), nil)$

\section{Don't-Know Constants}

In this section, we additionally propose a new class of constants called {\it don't-know constants}.
This class of constants is dual of
don't-care constants (a.k.a don't-care variables) and is not introduced properly in the literature.
To see why we need this class, consider the following database which contains the phone numbers of
the employees. \\\\
module emp. \\
phone(tom,434433) \\
phone(pete,200312) \\
phone(sue,*) \\
phone(john,*) \\
phone(tim,*) \\

\noindent In the above, * represents the don't-know constants in
that we do not know the phone numbers of
sue, john and tim.
Now suppose that sue and john share the same number for some reason.
 In this case, how do we represent this?  Representing this turns out to be tricky and a
reasonable solution is to use $\some x$ as shown below: \\\\
module emp. \\
$\some x \some y$ \\
phone(tom,434433) \\
phone(pete,200312) \\
phone(sue,x) \\
phone(john,x) \\
phone(tim,y) \\

\noindent It is worth noting that the class of don't-know constants is closely related to the class of
private/local constants. Over the years, the notion of private/local constants has been
a central concept in information hiding. However, we believe that this concept is rather unnatural
and the former class is a more fundamental and natural
 notion than the latter in information hiding and the latter can be derived from the former.

 Processing these prefixes within a module is not too difficult. In the initialization phase,
 the machine
needs to replace $\some x D$ with $D[\alpha/x]$ where $D$ is a module and $\alpha$ is a new Greek
letter. Greek symbols are used to denote that these constants are in fact $unknown$ to us.
For example, the above will be changed to the following during execution. \\\\
module emp. \\
phone(tom,434433) \\
phone(pete,200312) \\
phone(sue,$\alpha_1$) \\
phone(john,$\alpha_1$) \\
phone(tim,$\alpha_2$) \\

\section{Conclusion}\label{sec:conc}

In this paper, we have considered an extension to Prolog\cite{Bratko} with new noisy 
quantifiers.  This extension makes Prolog programs more versatile.
If we allow $\some x, \bsome x$ in $D$-formulas (and $\all x, \ball x$ in $G$-formulas), 
some important features -- information hiding \cite{MNPS91} and user interaction \cite{Jap03}, etc --
can be achieved. However, implementation becomes way more complicated.

 In the near future, we plan 
to investigate this possibility of including these features into logic 
programming.

\bibliographystyle{plain}


\end{document}

%% file: macros.tex
\newenvironment{numberedlist}
{\begin{list}{\makebox[20pt]{\hss(\arabic{itemno})\enspace}}
             {\usecounter{itemno}\labelwidth 20pt}}{\end{list}}

\newcounter{itemno}

\newcounter{itemno1}

\newcounter{itemno2}

\newcounter{exno}

\newcounter{defno}

\newenvironment{defn}{\refstepcounter{defno}\medskip \noindent {\bf
Definition \thedefno.\ }}{\medskip}

\newcommand{\sep}{\;\vert\;}

\newcommand{\oprove}{\vdash\kern-.6em\lower.7ex\hbox{$\scriptstyle O$}\,}

\newcommand{\pderivation}{{\cal P}\kern -.1em\hbox{\rm -derivation}}
\newcommand{\pderivationl}{{\cal P}\kern -.1em\hbox{\em -derivation}}
\newcommand{\pderivable}{{\cal P}\kern -.1em\hbox{\rm -derivable}}
\newcommand{\pderivablel}{{\cal P}\kern -.1em\hbox{\em -derivable}}
\newcommand{\pderivations}{{\cal P}\kern -.1em\hbox{\rm -derivations}}
\newcommand{\pderivability}{{\cal P}\kern -.1em\hbox{\rm -derivability}}

\newcommand{\all}{\forall}
\newcommand{\some}{\exists}

\newsavebox{\lpartfig}
\newsavebox{\rpartfig}

\newenvironment{exmple}{
 \begingroup \begin{tabbing} \hspace{2em}\= \hspace{3em}\= \hspace{3em}\=
\hspace{3em}\= \hspace{3em}\= \hspace{3em}\= \kill}{
 \end{tabbing}\endgroup}

\newcommand{\lb}{\langle}
\newcommand{\rb}{\rangle}

\newcommand{\intp}{intp_o}
  
     {\\* \hspace*{\fill} \end{trivlist}}

%% file: anonyi.bbl
\begin{thebibliography}{1}




\bibitem{Bratko}
I.~Bratko,   ``Prolog:programming for AI '',
 Addison Wesley, 2001 (3rd edition). 




\bibitem{Jap03}
G.~Japaridze, ``Introduction to computability logic'', Annals  of Pure and
 Applied  Logic, vol.123, pp.1--99, 2003.



\bibitem{KK07}
E.~Komendantskaya and V.~Komendantsky, ``On uniform proof-theoretical operational semantics for logic programming'',  In J.-Y. Beziau and A.Costa-Leite, editors, Perspectives on Universal Logic, pages 379--394. Polimetrica Publisher, 2007.

\bibitem{Mil89jlp}
D.~Miller, ``A logical analysis of modules in logic programming'', Journal of
  Logic Programming, vol.6, pp.79--108, 1989.

\bibitem{MNPS91}
D.~Miller, G.~Nadathur, F.~Pfenning, and A.~Scedrov, ``Uniform proofs as a
  foundation for logic programming'', Annals of Pure and Applied Logic, vol.51,
  pp.125--157, 1991.

 








\end{thebibliography}
